\documentclass[aps,prd,twocolumn,nofootinbib]{revtex4-1}

\usepackage{amssymb}
\usepackage{amsmath}
\usepackage{graphicx}
\usepackage[margin=5pt, font=small,labelfont=bf,justification=raggedright]{caption}

\usepackage{url}
\usepackage[bookmarks, pagebackref=false]{hyperref}
\usepackage{color}
	\definecolor{rossoCP3}{cmyk}{0,.88,.77,.40}
		\definecolor{graa}{rgb}{0.8,0.8,0.8}
		\definecolor{blaa}{rgb}{0.2,0.2,0.6}
		\definecolor{gron}{RGB}{0,150,0}
		\hypersetup{
			colorlinks, 
			bookmarksopen, 
			bookmarksnumbered,
			citecolor=blaa, 		
			linkcolor=black, 
			urlcolor=blaa,			
			}

\newcommand{\ea}[1]{%
\begin{align}
#1
\end{align}
}
\newcommand{\easl}[2]{%
\begin{equation}
\label{#1}
\begin{split}
#2
\end{split}
\end{equation}
}

\newcommand{\seal}[2]{%
\begin{subequations}
\label{#1}
\begin{align}
#2
\end{align}
\end{subequations}
}

\reversemarginpar

\def\Dil{\mathcal{D}}
\def\smallsum{\textstyle\sum\limits}

\begin{document}

\null\hfill\hbox{NORDITA-2015-134}\\

\vskip .3cm

\title{\Large 
New soft theorems for the gravity dilaton and the
\\Nambu-Goldstone dilaton at subsubleading order}

\author{
Paolo Di Vecchia$^{a,b}$,
Raffaele Marotta$^{c}$,
Matin Mojaza$^{b}$,
Josh Nohle$^{b}$
}

\affiliation{
\vskip .5cm
$^a$ The Niels Bohr Institute, University of Copenhagen,
\\ Blegdamsvej 17, DK-2100 Copenhagen \O , Denmark \\
\vskip .1mm
$^b$ NORDITA, KTH Royal Institute of Technology and Stockholm 
University,
\\Roslagstullsbacken 23, SE-10691 Stockholm, Sweden \\
\vskip .1mm
 $^c$  Istituto Nazionale di Fisica Nucleare, Sezione di Napoli,
\\ Complesso Universitario di Monte S. Angelo ed. 6, via Cintia, 80126, Napoli, Italy 
}

\begin{abstract}
We study the soft behavior of two seemingly different particles that are both referred to as dilatons in the literature, namely the one that appears in theories of gravity and in string theory and the Nambu-Goldstone boson of spontaneously broken conformal invariance. 
Our primary result is the discovery of a soft theorem at subsubleading order for each dilaton,
which in both cases contains the operator of special conformal transformations.
Interesting similarities as well as differences between the dilaton soft theorems are discussed.
\end{abstract}

\maketitle

\section{Introduction}

The word ``dilaton'' is currently used in particle physics for two different particles
that, \textit{a priori}, seem unrelated. One is the dilaton that appears in 
(super)gravity and string theories. This dilaton is a massless scalar particle accompanied by the graviton
and the Kalb-Ramond antisymmetric tensor in the massless sector 
of the closed bosonic string and in the NS-NS (Neveu-Schwarz) sector of the closed
superstring.  We refer to this particle as the ``gravity dilaton''.
The other dilaton is a Nambu-Goldstone (NG) boson arising from the
spontaneous breaking of scale and conformal invariance, which is frequently encountered 
in many phenomenological scenarios, for example in beyond-the-standard-model physics and inflationary cosmology. 
We refer to this as the ``NG dilaton''.
Although they are different particles, they share the feature of satisfying soft theorems,
according  to which the scattering of a low-energy particle is entirely determined by symmetry properties.
We will demonstrate that both dilatons obey soft  theorems through subsubleading
order in the low-energy expansion, i.e., through $\mathcal{O}(q^{1})$ in the soft-dilaton
momentum $q$.

The leading and subleading soft behavior of the gravity  dilaton has been  known since the  
seventies~\cite{Ademollo:1975pf,Shapiro:1975cz}. Furthermore, recent 
work has shown the leading and subleading behavior
to be a direct  consequence of the same gauge invariance 
that reveals the soft theorems for the  graviton~\cite{DiVecchia:2015oba}.
In this work, we extend this analysis to subsubleading order, revealing the role
of special conformal transformations.
In the case of a NG dilaton, the leading soft behavior vanishes for massless 
external states, while 
the subleading soft behavior is known to be a consequence of the Ward identity
of broken scale invariance~\cite{Callan:1970yg}. 

The similarity concerning  the soft behaviors of the two dilatons is discussed in a
recent paper by Boels and Wormsbecher~\cite{Boels:2015pta}, where it was explicitly shown that the term of $\mathcal{O}(q^0)$ in the soft-dilaton momentum $q$ is fixed from the Ward identity of scale invariance. It was furthermore conjectured that the term of $\mathcal{O}(q^1)$ is fixed from the Ward identities of special conformal transformations. 
In this paper, we explicitly derive the subsubleading soft behavior through the Ward identities of special conformal transformations, showing that the conjecture is correct.
Very recently the soft-theorem of the NG dilaton appearing in the proof of the a-theorem~\cite{Komargodski:2011vj,Elvang:2012st,Elvang:2012yc} has also been derived through $\mathcal{O}(q^0)$ in Ref.~\cite{Huang:2015sla}.
Furthermore, in Ref.~\cite{Larkoski:2014hta} 
the subleading soft terms in four-dimensional massless  gauge theories
have also been derived from conformal invariance.

The paper is organized as follows. In Sect.~\ref{string}, we demonstrate that the low-energy behavior
of the gravity dilaton through $\mathcal{O}(q^1)$ is entirely determined by gauge invariance
when the other (hard) particles are all either (massive) scalars or all gravitons or dilatons. 
In Sect.~\ref{Goldstone},  we show that the Ward identities 
of broken scale and conformal invariance completely determine the soft behavior 
of the NG dilaton through $\mathcal{O}(q^1)$.
Finally, in Sect.~\ref{conclusions} we compare the two behaviors and 
provide some conclusions.
We note that our derivation for the gravity dilaton soft theorem is valid at tree level.
For the NG dilaton, the formal arguments in our derivation hold at tree level in classically conformal theories. 
We also expect similar behavior at the quantum level in theories without a conformal anomaly.

\section{Gravity dilaton}\label{string}

We consider the low-energy behavior of the  dilaton that appears
in theories of (super)gravity and in string theories. The behavior in either case can be obtained 
using two different methods that, apart 
from possible string corrections, turn out to give the same result.

The first method that we employ is completely independent of string theory.
It consists of determining  the low-energy behavior of a tensor 
$M_{\mu \nu} (q; k_i)$ that, when saturated
with the polarization $\epsilon_{\mu \nu}$ of the graviton or of the dilaton,
describes the scattering amplitude of a graviton or a dilaton with momentum $q$
and $n$ other particles with momenta $k_i$. In this case, the soft behavior through
$\mathcal{O}(q^1)$ of  $M_{\mu \nu} (q; k_i)$ is fixed by imposing the 
following conditions,
\begin{eqnarray}
q^\mu M_{\mu \nu} (q; k_i ) = q^\nu M_{\mu \nu} (q; k_i ) =0\,,
\label{qMqMbis}
\end{eqnarray}
dictated by gauge invariance. The procedure is the one discussed in Ref.~\cite{BDDN}
without restricting oneself to the the assumption that the polarization of the soft particle is
traceless as in the case of the graviton. Proceeding in this way, when the other 
particles are (massive) scalars, i.e. not dilatons,
one gets the following soft behavior:
\ea{
& M^{\mu \nu} (q; k_i ) =    
\kappa_D  \sum_{i=1}^{n} \left[ 
 \frac{k_{i}^{\mu} k_{i}^{\nu}}{k_i \cdot q}- 
 i \frac{k_{i}^{\mu} q_\rho L_i^{\nu \rho} }{2 k_i \cdot q} 
  - i \frac{k_{i}^{\nu} q_\rho L_i^{\mu \rho} }{2 k_i\cdot q}  \right. 
 \nonumber \\
&  - \frac{1}{2} \frac{q_{\rho} L_i^{\mu 
\rho} q_{\sigma} L^{\nu \sigma}_i  }{k_i\cdot q}  
+ \left. \frac{1}{2} \left(  \eta^{\mu \nu}  q^\sigma - q^\mu \eta^{\nu \sigma} - 
\frac{k_{i}^{\mu} q^\nu q^\sigma  }{ k_i\cdot q} \right) \frac{\partial}{\partial k_i^{\sigma}} 
\right] 
\nonumber \\
&\times \mathcal{T}_n (k_{1},\ldots,k_{n})  + O(q^2) \ ,
\label{1gra4tacbvw}
}
where $\kappa_D$ is related to the $D$-dimensional 
Newton's constant by $\kappa_D = \sqrt{8\pi G_N^{(D)}}$,
$\mathcal{T}_n (k_{1},\ldots,k_{n})$ is the scattering amplitude of $n$ scalar particles
and 
\begin{eqnarray}
 L_i^{\mu \nu} = i \left( k_{i}^{\mu} \frac{
\partial}{\partial k_{i\nu}} - k_{i}^{\nu} \frac{
\partial}{\partial k_{i\mu}} \right)\,.
\label{Limunu}
\end{eqnarray}

When instead the other particles are gravitons and/or dilatons, one gets an extra polarization-dependent piece:
\ea{
&M^{\mu \nu}  (q ; k_i ) = \kappa_D
\sum_{i=1}^{n} \left[ 
\frac{k_{i}^{\mu} k_{i}^{\nu}}{k_i \cdot q}- 
 i \frac{k_{i}^{\mu} q_\rho J_i^{\nu \rho} }{2 k_i \cdot q} 
  - i \frac{k_{i}^{\nu} q_\rho J_i^{\mu \rho} }{2 k_i\cdot q}
 \right . \nonumber \\
& 
 - \frac{1}{2}   
\frac{q_\rho J_i^{\mu \rho} q_\sigma J_i^{\nu \sigma}}{k_i \cdot q }  
+ \frac{1}{2} \left(  \eta^{\mu \nu}  q^\sigma - q^\mu \eta^{\nu \sigma} - 
\frac{k_{i}^{\mu} q^\nu q^\sigma  }{ k_i\cdot q} \right) \frac{\partial}{\partial k_i^{\sigma}} 
\nonumber \\
& 
\left .
- \frac{1}{2} 
\frac{ q_\rho q_\sigma \eta_{\mu \nu} - q_\sigma q_\nu \eta_{\rho \mu} 
- q_\rho q_\mu \eta_{\sigma \nu}}{k_i \cdot q} \epsilon_{i}^{\rho} 
\frac{\partial}{\partial \epsilon_{i\sigma}} \right] \nonumber \\
&
\times M_n ( k_1,\ldots,k_n )
 + {\cal{O}} (q^2 )
 \,, 
\label{S0+S1+S2hhh}
}
where $M_n( k_1,\ldots,k_n )$ is the $n$-point scattering amplitude involving  gravitons and/or dilatons and,
\easl{Sibbb}{
(J_i ) _{\mu \nu} &= (L_i)_{\mu \nu} + (S_i)_{\mu \nu} \,,
 \\[.2cm]
(S_i )_{\mu \nu} &= i \left( \epsilon_{i \mu} \frac{\partial}{
\partial\epsilon_{i}^{\nu}} -
\epsilon_{i \nu} \frac{\partial}{\partial\epsilon_{i}^{\mu}} \right) \,.
}

It is easy to check that both tensors $M_{\mu \nu} (q ; k_i )$ of Eqs.~\eqref{1gra4tacbvw} and \eqref{S0+S1+S2hhh} satisfy Eq.~\eqref{qMqMbis}.
By saturating the two previous tensors with the polarization of 
the dilaton, given by
\ea{
\epsilon^{\mu \nu}_d = (\eta^{\mu \nu} - q^\mu {\bar{q}}^\nu - q^\nu {\bar{q}}^\mu)
/\sqrt{D-2}\, ,} 
where $q^2 = {\bar{q}}^2 =0$ and
$q\cdot{\bar{q}} =1$,
we get the soft behavior of a dilaton. The amplitude of a soft dilaton and $n$ scalars then reads:
\ea{
& \epsilon^{\mu \nu}_d M_{\mu \nu} (q; k_i)  = \frac{\kappa_D}{\sqrt{D-2}}
 \Big \{  - \sum_{i=1}^{n} \frac{ m_i^2 }{k_i\cdot q}
\nonumber\\
&  
 \times \left( 1 + q^{\rho}  \frac{\partial}{\partial k_{i}^{\rho}} + \frac{q^{\rho}   q^{\sigma}}{2}  \frac{ \partial^2}{ \partial k_{i}^{\rho} 
\partial k_{ i}^{\sigma}   }  \right) 
+ 2 -    \sum_{i=1}^{n} k_{i}^{\mu}   \frac{\partial}{ \partial k_{i}^{\mu }}
\nonumber\\
&  
 + \frac{q^{\rho}}{2} \sum_{i=1}^{n}\left(2\,k_{i}^{\mu}\frac{\partial^2}{\partial k_{i}^{\mu} \partial k_{i}^{\rho}} 
 - k_{i\rho} \frac{\partial^2}{\partial k_{i}^{\mu} \partial k_{i\mu}} \right)
  \Big \}
 \mathcal{T}_n + \mathcal{O}(q^{2})\, ,
\label{dilantacx}   
}
where $m_i$ is the mass of the $i^{\text{th}}$ scalar particle.

Similarly, by saturating Eq.~(\ref{S0+S1+S2hhh}) with the dilaton polarization, one 
gets the soft behavior
of a dilaton in an amplitude with hard gravitons and/or dilatons: 
\ea{
&\epsilon^{\mu \nu}_{d}  M_{\mu \nu}(q; k_{i})  =\frac{\kappa_D}{\sqrt{D-2}} 
\Big\{  
2 - \sum_{i=1}^{n}k_{i}^{\mu}   \frac{\partial}{ \partial k_{i}^{\mu }}
\nonumber \\
& 
 + \frac{q^{\rho}}{2} \sum_{i=1}^{n}k_{i}^{\mu} \left [\left(2\,k_{i}^{\mu}\frac{\partial^2}{\partial k_{i}^{\mu} \partial k_{i}^{\rho}} 
 - k_{i\rho} \frac{\partial^2}{\partial k_{i}^{\mu} \partial k_{i\mu}} \right) 
 -i q^\rho S^{(i)}_{\mu \rho} \frac{\partial}{\partial k_{i\mu}} 
\right] 
\nonumber \\
&  
+ \sum_{i=1}^{n} \frac{q^\rho q^\sigma}{ 2 k_i\cdot q} \left(      
( S_{ \rho \mu  }^{(i)} )   \eta^{\mu \nu}  (S_{\nu \sigma }^{(i)} ) 
  +   D     \epsilon_{i\rho} 
\frac{\partial}{\partial \epsilon_{i }^{\sigma}} \right)
 \Big\}M_{n}   + {\cal{O}} (q^2 ) \, .
\label{M(n+1)xyyw}
}

In all of our expressions, whenever a momentum derivative is acting on 
an amplitude, it is implicitly assumed that momentum conservation 
is applied to one of the external momenta in the amplitude.

The second method is to consider a string theory,
compute an amplitude with a dilaton and
study its behavior when the momentum of the dilaton is soft. 
(For recent studies of soft theorems in string theory see~\cite{Schwab:2014fia,Bianchi:2014gla,Schwab:2014sla,Bianchi:2015yta,
Volovich:2015yoa,DiVecchia:2015bfa,Guerrieri:2015eea,Bianchi:2015lnw}.)
It turns 
out that, if the other particles
are tachyons of the bosonic string, one gets exactly the behavior in 
Eq.~(\ref{dilantacx}) with
$m_i^2 = - \frac{4}{\alpha'}$, as shown in Ref.~\cite{DiVecchia:2015oba}. 
On the other hand, if the other particles are also dilatons and/or
gravitons, one finds in Refs.~\cite{Ademollo:1975pf,DiVecchia:2015oba,tobepub} the behavior given in Eq.~(\ref{M(n+1)xyyw}).

\section{Dilaton of  broken conformal invariance}\label{Goldstone}

We consider a field theory whose  action is invariant under some 
transformation, with corresponding Noether current $j^{\mu}$, and we study the matrix element
$
T^*\langle 0| j^{\mu} (x) \phi (x_1 )   \dots \phi (x_n) | 0\rangle \,,
$
 where $T^*$ denotes the $T$-product with the derivatives placed 
outside of the time-ordering symbol. For the sake of simplicity, here and in the following
we restrict ourselves to scalar fields. Taking  the derivative of this quantity with respect to the variable $x$
and subsequently performing a Fourier transformation in $x$
yields a form of the Ward identity that we use to derive low-energy theorems,
\easl{LowEnergyWard}{
&
\int d^{D}x\,\text{e}^{-i q\cdot x}
\Bigl [
-\partial_\mu \,T^*\langle 0| j^{\mu} (x) \phi (x_1 ) \dots \phi (x_n) | 0\rangle
 \\
&  \hspace{2.2cm}
+
T^*\langle 0| \partial_\mu  j^{\mu} (x) \phi (x_1)   \dots \phi (x_n) | 0\rangle
\Bigr ] \\ &
=
-\sum_{i=1}^{n}\,\text{e}^{-i q\cdot x_{i}}
\,T^{*}\langle 0| \phi (x_1) \dots   \delta \phi (x_i)\dots \phi (x_n)| 0 \rangle\,,
}
where $\delta \phi$ is the infinitesimal transformation of the field $\phi$ under
the generators  of the symmetry.
We consider the case where the Noether current   corresponds
to  scale and special conformal transformations of a spontaneously broken
conformal field  theory in $D$ space-time dimensions.%
\footnote{For a review on 
the conformal Ward identities, see  Coleman's beautiful book~\cite{Coleman}.}

Let us start by discussing
the scale transformation. The Noether current and its divergence are equal to,
\begin{eqnarray}
j^\mu_\Dil = x_\nu T^{\mu \nu}  ~~;~~ \partial_{\mu}\,j_{\Dil}^{\mu} = T^{\phantom{\mu}\mu }_{\mu} \, ,
\label{DCurrent}
\end{eqnarray}
where  $T^{\mu \nu}$ is the energy-momentum tensor of the theory. The  action of 
the generator of scale transformation $\Dil$ on a scalar field is  given by:
\begin{eqnarray}
\delta \phi (x ) = [ \Dil , \phi (x) ]= i 
 \left( d + 
 x^\mu \partial_\mu 
\right) \phi (x)\, ,
\label{QD}
\end{eqnarray}
where   $d$ is the scaling dimension of the 
scalar field $\phi (x)$. 

Considering the left-hand side of Eq.~(\ref{LowEnergyWard}),
we may neglect the first term by only keeping terms up to $\mathcal{O}(q^0)$ in the soft expansion,
assuming that 
$T^*\langle 0| j^{\mu} (x) \phi (x_1 ) \dots \phi (x_n) | 0\rangle$
does not have 
a pole at \mbox{$q=0$}. 
In a theory with spontaneously broken conformal symmetry, the
second term contributes. 
Notably, in such a theory the NG dilaton $\xi(x)$, which
is the massless fluctuation around the nonconformal vacuum,
is by its equation of motion related to the
the trace of the energy-momentum tensor in 
the following way:
\begin{eqnarray}
T_{\mu}^{\phantom{\mu}\mu}  (x) = - {v} \, \partial^2\,\xi (x)\,,
\label{Tmumud}
\end{eqnarray}
where $v$ is related to the vacuum expectation value 
of the dilaton field, denoted by $\langle \xi \rangle$.
(In the specific  theory considered in 
Ref.~\cite{Boels:2015pta}, we find that ${v}= \frac{D-2}{2} 
\langle \xi \rangle$, where  Eq. (\ref{Tmumud}) is a consequence of 
the classical equations of 
motion. See also Sect.~2 of Ref.~\cite{Schwimmer:2010za} and Sect.~3 of Ref.~\cite{Higashijima:1994}.) 
It follows from Eq.~(\ref{DCurrent}) that
\begin{align}
\partial_{\mu}\,j_{\Dil}^{\mu}(x) = {v}\, (-\partial^2)\,\xi (x)\,.
\label{Ttrace}
\end{align}

To translate the correlation function identity in Eq.~(\ref{LowEnergyWard}) to an identity among amplitudes, we apply the LSZ reduction. We define the LSZ operator,
\begin{align}
\Bigl[\text{LSZ}\Bigr] 
\equiv
i^{n}\left(\prod_{j=1}^{n}\lim_{k_{j}^{2}\rightarrow -m_{j}^{2}}\int d^{D}x_{j}\,\text{e}^{-i k_{j}\cdot x_{j}}(-\partial_{j}^{2}+m_{j}^{2})\right)\,,
\nonumber 
\end{align}
where the limits $k_j^2 \to - m_j^2$ put the external states 
on shell, which has to be performed only at the end.
Inserting Eq.~\eqref{Ttrace} into Eq.~(\ref{LowEnergyWard}) and 
applying the LSZ reduction
 we find the left-hand side 
of Eq.~(\ref{LowEnergyWard}) to yield,
\begin{align}
&\Bigr[\text{LSZ}\Bigr]
\int d^{D}x\,\text{e}^{-i q\cdot x}
\,T^*\langle 0|\partial_\mu j_{\Dil}^{\mu} (x) \phi (x_1)   \dots \phi (x_n) | 0\rangle
\notag \\ &
=
(-i)\,{v}\,(2\pi)^{D}
{\delta^{(D)}(\smallsum_{j=1}^{n}k_{j}+q)}
\mathcal{T}_{n+1}(q; k_{1},\ldots,k_{n})\,,
\label{DilLHS}
\end{align}
where we have Fourier transformed and extracted the poles of the correlation 
function to identify the amplitude $\mathcal{T}_{n+1}$.%
\footnote{For more details, see for instance Ref.~\cite{Srednicki}, whose 
conventions we follow up to an immaterial factor $i$.
}
Notice that the operator $(-\partial^{2})$ from the divergence of the current effectively amputates the dilaton propagator $\langle \xi \xi \rangle \sim 1/q^{2}$.
It is implicitly assumed that this expression holds only up to terms of $\mathcal{O}(q^0)$,
and that the $k_{i}$ on-shell limits are taken after the soft expansion. 

Next, performing the same operations on the right-hand side of Eq.~(\ref{LowEnergyWard}) gives,
\begin{align}
&\Bigl[\text{LSZ}\Bigr]
\left(
-
\sum_{i=1}^{n}\text{e}^{-i q\cdot x_{i}}\,
T^{*}\left<0\right|\phi(x_{1})\cdots\delta\phi(x_{i})\cdots\phi(x_{n})\left|0\right>
\right)
\notag \\ &
=
-
\sum_{i=1}^{n} \Biggl[
\lim_{k^{2}_{i}\rightarrow -m_{i}^{2}}(k_{i}^{2}+m_{i}^{2})
\,i\left(d-D-(k_{i}+q)^{\mu}\frac{\partial}{\partial k_{i}^{\mu}}\right)
\notag \\ &\hspace{0cm}
\times 
\frac{(2\pi)^{D}\delta^{(D)}\left(\sum_{j=1}^{n}k_{j}+q\right)}
{
(k_{i}+q)^{2}+m_{i}^{2}
} \mathcal{T}_{n}(k_{1},\ldots,k_{i}+q,\ldots,k_{n})
\Biggr]
\,,
\label{LSZ1}
\end{align}
where all states $j\neq i$ have already been amputated and put on shell.
The next step is to commute the differential operator past the $i^{\text{th}}$ propagator and the $\delta$-function, using the identity:
\begin{align}
&\sum_{i=1}^{n}k^{\mu}_{i}\frac{\partial}{\partial k_{i\,\nu}}\left[\delta^{(D)}
(\smallsum_{j=1}^{n}k_{j})
\mathcal{T}_{n}(k_{1},\ldots,k_{n})\right]
=
\delta^{(D)}
(\smallsum_{j=1}^{n}k_{j})
\notag \\ &\hspace{.3cm}
\times
\left(-\eta^{\mu\nu}+\sum_{i=1}^{n}k^{\mu}_{i}\frac{\partial}{\partial k_{i\,\nu}}\right)
\mathcal{T}_{n}(k_{1},\ldots,-\smallsum_{j=1}^{n-1}k_{j})\,.
\label{deltaID}
\end{align}
To ensure a simple form, notice the need to apply momentum conservation on one of the momenta of the $n$-point amplitude before differentiating. As we remarked in the previous section, it is necessary to enforce this condition whenever a derivative is acting on the amplitude. We denote this procedure for brevity by $\bar{k}_n = - \sum_{j=1}^{n-1}k_{j}$ 
(see also Ref.~\cite{Broedel:2014fsa} for a more general discussion).
Expanding $\mathcal{T}_n$ in the soft momentum $q$ and following through with this procedure, we find
Eq.~\eqref{LSZ1}, up to terms of $\mathcal{O}(q^1)$, to be equal to (the $\delta$-function is kept unexpanded because it appears in the same form on the left-hand side):
\begin{align}
&-i(2\pi)^{D}\delta^{(D)}({\smallsum_{j=1}^{n}k_{j}+q})
\Bigg \{
D - n\,d - \sum_{i=1}^{n}k_{i}^{\mu}\frac{\partial}{\partial k_{i}^{\mu}}
\nonumber \\
&\ \ \
- \sum_{i=1}^n \lim_{k_i^2 \to -m_i^2} \frac{2m_i^2 (k_i^2 + m_i^2)}{\left [(k_i+q)^2 + m_i^2\right ]^2} 
\left ( 1 + q^\mu \frac{\partial}{\partial k_i^\mu} \right)
\Bigg\} 
\nonumber \\
& \ \ \ \times
\mathcal{T}_{n}(k_{1},\ldots,\bar{k}_n )\,,
\end{align}
where 
we have used $d = (D-2)/2$, and neglected terms of $\mathcal{O}(q^1)$.
The second line clearly has a singularity problem, and it would be incorrect to expand
the denominator in the soft momentum $q$, since the on-shell limit is then divergent.
The soft momentum here must instead be understood as a physical regulator that
leaves the on-shell limit finite, which physically means that we should identify $k_i \simeq k_i + q$,
leaving us with
\easl{DilRHS}{
&-i(2\pi)^{D}\delta^{(D)}({\smallsum_{j=1}^{n}k_{j}+q})
\Bigg \{
D - n\,d - \sum_{i=1}^{n}k_{i}^{\mu}\frac{\partial}{\partial k_{i}^{\mu}}
 \\
&\ \ \
- \sum_{i=1}^n \frac{m_i^2}{k_i \cdot q} 
\left ( 1 + q^\mu \frac{\partial}{\partial k_i^\mu} \right)
\Bigg\} 
\mathcal{T}_{n}(k_{1},\ldots,\bar{k}_n )\,.
}

As we will see, this regulation leads to physically sensible results when one
considers the decomposition of $\mathcal{T}_{n+1}$ as,
\easl{Texpansion}{
\mathcal{T}_{n+1}
=& 
\left [
\sum_{i=1}^n\frac{\mathbf{S}_i^{(-1)}(q)}{k_{i}\cdot q}
+ \mathbf{S}^{(0)} + q^\mu \mathbf{S}^{(1)}_\mu
\right ] 
 \\
&
\times \mathcal{T}_n(k_1, \ldots, k_n)
+ \mathcal{O}(q^2) \, ,
}
where the $\mathbf{S}^{(0)}$ and $\mathbf{S}_\mu^{(1)}$ are operators dependent only on the momenta $k_{j}$, while $\mathbf{S}^{(-1)}$ may depend on $q$ as well.
Then, from equating Eqs.~(\ref{DilLHS})~and~(\ref{DilRHS}), we find,
\seal{DFinal}{
& v \mathbf{S}_i^{(-1)} =
- m_i^2 \left ( 1 + q^\mu \frac{\partial}{\partial k_i^\mu} \right)
+ \mathcal{O}(q^2)\,,
 \\
&v\mathbf{S}^{(0)} =
D - n\,d - \sum_{i=1}^{n}k_{i}^{\mu}\frac{\partial}{\partial k_{i}^{\mu}}\,.
 }
The previous expressions agree with Eq.~(4) of Ref.~\cite{Callan:1970yg} for the four-dimensional massless case.
One can see that the Ward identity for the scale transformation completely 
determines the $\mathcal{O}(q^{-1})$ and $\mathcal{O}(q^0)$ contributions of the amplitude with a soft dilaton 
in terms of the amplitude without the dilaton.

We now show that the Ward identity of the special conformal transformation~\cite{Parisi:1972zy} determines also the $\mathcal{O}(q^1)$ contribution of the amplitude with a soft dilaton. In this case, the Noether
current and its divergence are given by,
\easl{Noetheconf1}{
j^\mu_{\phantom{\mu}(\lambda)} &= T^{\mu \nu} ( 2 x_\nu x_\lambda - 
\eta_{\nu \lambda} x^2 ) \ , \\
\partial_\mu\,j^\mu_{\phantom{\mu}(\lambda)} &= 2\,x_{\lambda}\,T^{\mu}_{\phantom{\mu}\mu}
=2\,{v}\,x_{\lambda}(-\partial^2)\,\xi (x)
\,,
}
while the action of a special conformal transformation on a scalar field is equal to,
\easl{Klambda1}{
\delta_{(\lambda)} \phi (x) 
&= \left[\mathcal{K}_{\lambda},\phi(x)\right]  \\
&= i
\left( (2 x_\lambda x_{\nu} - \eta_{\lambda \nu} x^2 ) \partial^\nu 
+ 2\,d\,x_\lambda \right)\phi(x)\,.
}
(Note that, in general, the special conformal transformation has an extra term when acting on fields with spin.)

Analyzing the right-hand side of Eq.~(\ref{LowEnergyWard}),
mirroring the procedure for scale transformations utilized above,
 we find an expression analogous to Eq.~(\ref{DilRHS}):
\begin{align}
&\bigl[\text{\small LSZ}\bigr]
\Bigl(
{-\smallsum_{i=1}^{n}}
\text{e}^{-i q\cdot x_{i}}
T^{*}\left<0\right| \phi(x_{1})\cdots\delta_{(\lambda)}\phi(x_{i})\cdots\phi(x_{n})\left|0\right>
\Bigr)
\notag \\ &
=
(2\pi)^{D}\delta^{(D)}(\smallsum_{j=1}^{n}k_{j}+q)
\times 
2{\smallsum_{i=1}^{n}}
\Big\{ 
\nonumber \\
&\hspace{.5cm}
\frac{m_i^2}{k_{i}\cdot q} \left[ \frac{k_{i\,\lambda}}{ k_{i}\cdot q   } 
- \frac{\partial}{\partial k_{i}^{\lambda} }\right] 
 \left (1 + q^\mu\frac{\partial}{\partial k_i^\mu} + \frac{1}{2}q^\mu q^\nu \frac{\partial^2}{\partial k^\mu \partial k^\nu} \right )
\nonumber \\ 
& \hspace{.5cm}
-\left[ k_i^{\mu} \left(\frac{\partial^2}{\partial k_{i}^{\mu} 
\partial k_{i}^{\lambda}} -
\frac{1}{2} \eta_{\mu\lambda}  \frac{\partial^2}{\partial k_{i\nu} \partial k_{i}^{\nu}}  
 \right)
+ d \frac{\partial}{\partial k_{i}^{\lambda} } \right]
\notag \\ &\hspace{.5cm}  
+ \mathcal{O}(q) \Big\} 
\mathcal{T}_{n}(k_{1},\ldots,\bar{k}_n)\,,
\label{KRHS}
\end{align}
where again we  use $d = (D-2)/2$.
This calculation is very similar to that of scale transformations except that the action of the differential operators 
on the $\delta$-function does not introduce extra terms, like $D$ in the case of scale transformations.
Similarly, analyzing the left-hand side of Eq.~(\ref{LowEnergyWard}), we find an expression analogous to Eq.~(\ref{DilLHS}), but now with a derivative of $q$ acting on the amplitude:
\begin{align}
&\Bigr[\text{LSZ}\Bigr]
\int d^{D}x\,\text{e}^{-i q\cdot x}
\,T^*\langle 0|\partial_\mu j^{\mu}_{\phantom{\mu}(\lambda)} (x) \phi (x_1)   \dots \phi (x_n) | 0\rangle
\notag \\ &\hspace{.5cm}
=
2\,{v}\,(2\pi)^{D}\delta^{(D)}({\smallsum_{j=1}^{n}k_{j}}+q)
\notag \\[-3mm]
&\hspace{1cm}\times\frac{\partial}{\partial q^{\lambda}}
\mathcal{T}_{n+1}(q; k_{1},\ldots,-q-\smallsum_{j=1}^{n-1}k_{j})\,.
\label{KLHS}
\end{align}
Equating Eqs.~(\ref{KLHS})~and~(\ref{KRHS}) and contracting with $q^{\lambda}$, we find

\begin{align}
{v}\,
q^{\lambda}
&\frac{\partial}{\partial q^{\lambda}}
\mathcal{T}_{n+1}(q; k_{1},\ldots,-q-\smallsum_{j=1}^{n-1}k_{j})
\nonumber \\ 
 =&
\sum_{i=1}^{n}
\left\{
\frac{m_i^2}{k_{i}\cdot q} 
\left ( 1- \frac{1}{2} q^\mu q^\lambda \frac{\partial^2}{\partial k^\mu \partial k^\lambda} \right )
\right. 
\nonumber \\
&\left. 
-q^{\lambda}\left[ k_i^{\mu} \left(\frac{\partial^2}{\partial k_{i}^{\mu} 
\partial k_{i}^{\lambda}} -
\frac{1}{2} \eta_{\mu\lambda}  \frac{\partial^2}{\partial k_{i\nu} \partial k_{i}^{\nu}}  
 \right)
+ d \frac{\partial}{\partial k_{i}^{\lambda} } \right]
\right\} 
\notag \\ &
\times 
\mathcal{T}_{n}(k_{1},\ldots,\bar{k}_n)+ \mathcal{O}(q^2) \,.
\label{KFinal}
\end{align}

With Eq.~(\ref{DFinal}) in hand, we may use Eq.~\eqref{Texpansion} to replace $\mathcal{T}_{n+1}$.
The $\mathcal{O}(q^{-1})$ terms then exactly cancel on both sides of the above equation, leading to an
equation that uniquely determines the $\mathcal{O}(q^1)$ terms of $\mathcal{T}_{n+1}$, i.e.
\seal{orderq}{
& v \mathbf{S}_i^{(-1)}\Big |_{\mathcal{O}(q^2)} =
- m_i^2 \left (
\frac{1}{2}q^\mu q^\lambda \frac{\partial^2}{\partial k^\mu \partial k^\lambda} \right)\,,
 \\
&v \mathbf{S}_\lambda^{(1)} 
 =
\sum_{i=1}^n \left[ \frac{k_{i\lambda}}{2} \frac{\partial^2}{\partial k_{i\nu} \partial k_{i}^{\nu}} 
-k_i^{\mu}\frac{\partial^2}{\partial k_{i}^{\mu} 
\partial k_{i}^{\lambda}}
- d \frac{\partial}{\partial k_{i}^{\lambda} } \right]
}

In conclusion, the Ward identities of the scale and special conformal transformations determine
completely the low-energy behavior, through the $\mathcal{O}(q^1)$,  of an amplitude with a soft dilaton 
in terms of the amplitude without the dilaton. Inserting the results from Eqs.~(\ref{DFinal}) and \eqref{orderq} into Eq.~(\ref{Texpansion}), we have in total,
\begin{widetext}
\easl{finalbehavior}{
 {v}\,{\cal{T}}_{n+1} ( q; k_1, \ldots, k_n ) =
\Bigg\{ 
& - \sum_{i=1}^{n} \frac{m_i^2}{k_i \cdot q}
\left ( 1 + q^\mu \frac{\partial}{\partial k_i^{\mu}}
+ \frac{1}{2} q^\mu q^\nu \frac{\partial^2}{\partial k_i^\mu \partial k_i^{\nu} }\right)
+ D - nd - \sum_{i=1}^{n}k_i^{\mu} \frac{\partial}{\partial k^{\mu}_i}
 \\[.2cm] &
  \left.
-q^\lambda \sum_{i=1}^{n} \left[ 
\frac{1}{2}
\left(2\,k_i^{\mu} \frac{\partial^2}{\partial k_{i}^{\mu} 
\partial k_{i}^{\lambda}} -
k_{i\,\lambda}   \frac{\partial^2}{\partial k_{i\nu} \partial k_{i}^{\nu}}  
 \right)
+ d\,\frac{\partial}{\partial k_{i}^{\lambda} } \right] 
 \right\}
\mathcal{T}_{n}(k_{1},\ldots,\bar{k}_n) + {\cal{O}} (q^2 )\,.
}
\end{widetext}
Notice that the terms proportional to 
$m_i^2$ can be considered as the expansion of 
\mbox{$\mathcal{T}_n(k_1,\ldots, k_i+q, \ldots, k_n)$} in $q$. Indeed, this 
is exactly what we expect from the structure of tree-level amplitudes, 
however, here it comes out as a consequence of the Ward identities.
We have checked the above expression against 3-, 4-, 5- and 6-point 
amplitudes in a simple four-dimensional two-scalar model, and 
against 3-, 4-, and 5-point amplitudes in the generalized $D$-dimensional 
model, discussed in Ref.~\cite{Boels:2015pta}. The term of order 
$\mathcal{O}(q^0)$ agrees with the one proposed 
in Ref.~\cite{Huang:2015sla}.\footnote{We thank Congkao Wen for 
communicating to us that the dilaton actions constructed for proving the
a-theorem~\cite{Komargodski:2011vj,Elvang:2012st,Elvang:2012yc} satisfy the complete soft behavior in Eq.~(\ref{finalbehavior}).}

\section{Comparison and conclusions}
\label{conclusions}

In this paper, we have extracted the tree-level
soft behavior of two, \textit{a priori}, different objects 
that are both referred to as dilatons in the literature.
We have shown that in both cases the symmetry properties determine the soft
behavior through the $\mathcal{O}(q^1)$ in the dilaton momentum. In the case of the
gravity dilaton, the symmetry is the same gauge invariance that determines 
the soft behavior of the graviton, while the soft behavior of the
NG dilaton
is determined by the Ward identities
of scale and special conformal transformations. The soft behavior of the 
gravity dilaton is given in Eq.~(\ref{dilantacx}) in an amplitude with scalar particles 
and in Eq.~(\ref{M(n+1)xyyw}) in an amplitude with other massless particles, while that 
of a NG dilaton is given in Eq.~(\ref{finalbehavior}). 
In both cases, we get a term of $\mathcal{O}(q^{-1})$, which is proportional to the squared mass
of the other particles.  This  follows from the fact that, in both cases, 
there is a three-point amplitude involving a dilaton and two identical particles, which 
is proportional to their squared mass.  

Furthermore, for both dilatons
we have a term of $\mathcal{O}(q^0)$ and a term of 
$\mathcal{O}(q^1)$ that are fixed by the symmetry properties and which contain terms connected 
to the conformal operators $\Dil$ and $\mathcal{K}_\mu$. 
The generators of space-time scale transformations, $\hat{\Dil}$, and special conformal space-time transformation $\hat{\mathcal{K}}_\mu$, are equal to 
\seal{DKmu}{
\hat{\Dil} &= x_\mu \hat{\mathcal{P}}^\mu 
\, ,
 \\
 \hat{\mathcal{K}}_{\mu} &= (2 x_\mu x_{\lambda} - x^2 \eta_{\mu \lambda} ) 
\hat{\mathcal{P}}^\lambda
\, ,
}
where $\hat{\mathcal{P}^\mu}$ is the generator of space-time translations.
Going to momentum space they become:
\seal{DKmumom}{
\hat{\Dil} &= - i k_{\mu} \frac{ \partial}{\partial k_\mu} \, , \\
\hat{\mathcal{K}}_{\mu} &= - \left( 2k^{\nu} 
 \frac{\partial^2}{\partial k^\nu \partial k^\mu}-   k_\mu
 \frac{ \partial^2}{ \partial k^{\nu} \partial k_{\nu} } \right) \, ,
}
which are precisely the operators that appear in both soft behaviors.
Apart from these similarities, there seems to be some difference in the soft 
behavior as well;  
looking at the term of $\mathcal{O}(q^0)$ in Eqs.~\eqref{dilantacx} and \eqref{finalbehavior}, we find that in 
the first case the kinematically invariant part equals $2$, while in the second case it is equal to $D- nd$.   The term $D- nd = D - n \frac{D-2}{2}$ represents the fact that $\mathcal{T}_n$ has exactly this mass dimension. In fact, for dimensional reasons, the amplitude $\mathcal{T}_n$ 
has the following form:
\begin{eqnarray}
\mathcal{T}_n\left(m; k_i \right ) = m^{D - n \frac{D-2}{2}} g^{n-2} G_n \left( {k_i}/{m} \right)\,,
\label{Tndimensi}
\end{eqnarray}
where $m$ is the mass scale of the theory, which is typically 
given by $(g v)^{\frac{2}{D-2}}$, where $v$ is related to the vacuum expectation value 
of the dilaton field,
$\langle \xi \rangle$, and $g$ is a typical dimensionless coupling constant of the theory. 
It follows immediately from Eq.~\eqref{Tndimensi} that the term of $\mathcal{O}(q^0)$ in Eq.~\eqref{finalbehavior}  can be rewritten as:
\ea{
&\frac{1}{{v}} \left( D - n\frac{D-2}{2} - 
\sum_{i=1}^{n}  k_i^\mu \frac{\partial}{\partial k_{i}^\mu} \right) \mathcal{T}_n 
\nonumber \\
&\hspace{.5cm} = \frac{m}{{v}} \frac{\partial}{\partial m} \mathcal{T}_n
 = \left (\frac{D-2}{2}\right ) 
 \frac{\partial}{\partial v} \mathcal{T}_n 
 \sim \frac{\partial}{\partial \langle \xi \rangle} \mathcal{T}_n \, ,
\label{fiu}
}
where $\sim$ means up to a numerical constant.

The term of $\mathcal{O}(q^0)$ in Eq.~\eqref{dilantacx} instead seems to have another meaning. 
In string theory $M_n$ has, of
course, the same physical dimension as $\mathcal{T}_n$ and
the following form: 
\ea{
 M_n &= \frac{4\pi}{\alpha'} \left( \frac{\kappa_D}{\pi} \right)^{n-2} 
F_n \left( \sqrt{\alpha'} k_i \right)  
\nonumber \\
&=
C_n m_s^{D - n \frac{D-2}{2}} g_s^{n-2} 
F_n \left (  {k_i}/{m_s} \right) \, , 
\label{Mndimana}
}
where $\alpha'$ is the inverse string tension, and in the second line we rewrote the expression 
into a form similar to Eq.~\eqref{Tndimensi}, with
 $C_n$ being a numerical constant, $m_s \equiv {1}/{\sqrt{\alpha'}}$, $g_s$ the string coupling constant and
\begin{eqnarray}
\kappa_D = \frac{1}{2^{\frac{D-10}{4}} } 
\frac{g_s}{\sqrt{2}} (2\pi)^{
\frac{D-3}{2}} ( \sqrt{\alpha'})^{\frac{D-2}{2}}\,\, .
\label{mskappad}
\end{eqnarray}  
In the field theory limit (gravity and supergravity), $M_n$  behaves 
as follows:
\begin{eqnarray}
\lim_{\alpha'\rightarrow 0} M_n \sim  \left( \frac{\kappa_D}{\pi} \right)^{n-2} 
\lim_{\alpha'\rightarrow 0} \frac{ 4 \pi F_n (  \sqrt{\alpha' }k_i )}{\alpha'}\,,
\label{supergralimit}
\end{eqnarray}
where the limit is  finite  and,
for dimensional reasons, 
has
 to provide a homogenous function in  the momenta of the particles
  of degree $2$. This means that, in this 
limit, the action
of the full term of $\mathcal{O}(q^0)$ in Eqs.~\eqref{dilantacx}-\eqref{M(n+1)xyyw} gives zero.  In other words, in the field
theory  limit, the amplitude with one soft dilaton is  
vanishing
 and this is
consistent with the fact that an amplitude with an odd number of dilatons in (super)gravity  
vanishes.
However,  in the full string theory this is no longer true. 
Equation~(\ref{Mndimana}) 
implies that 
the $\mathcal{O}(q^0)$ term in Eqs.~\eqref{dilantacx}-\eqref{M(n+1)xyyw}
can be written as:
\begin{eqnarray}
&&\kappa_D \left (2 - \sum_{i=1}^{n}  k_i^\mu \frac{\partial}{\partial k_{i}^\mu} \right )
M_n 
\nonumber \\
&&\hspace{.5cm} =
\kappa_D \left( \frac{D-2}{2} g_s \frac{\partial}{\partial g_s} -
 \sqrt{\alpha'} 
\frac{\partial}{\partial\sqrt{\alpha'}} \right)M_n
\nonumber \\
&&\hspace{.5cm} =
\left (\frac{D-2}{2}\right )\kappa_D  \frac{d}{d \phi_0}
M_n\,,
\label{lintre}
\end{eqnarray}
where we used the relation between $g_s$ and the vacuum expectation value 
of the string dilaton, $\phi_0$, i.e. $g_s \equiv {\rm e}^{\phi_0}$. (Eq. (\ref{lintre}) is valid also when the amplitude $M_n$ contains massless open strings.)
The operator that appears in the second line
leaves $\kappa_D$ invariant, which can be explicitly checked (see also~\cite{Yoneya:1987gc,Hata:1992it}).
This implies that in string theory one does not have two 
fundamental constants $g_s \equiv {\rm e}^{\phi_0}$ and $\alpha'$ that 
can be  fixed  independently from each other; the physical amplitudes 
depend on $\alpha'$ and on $\kappa_D$ where a change of $\phi_0$ 
can be reabsorbed in a rescaling of $\alpha'$. 
Thus, the last step in Eq. (\ref{lintre}) means that we should differentiate with respect to $\phi_0$ keeping $\kappa_D$ fixed. To compare with the field theory dilaton, we should canonically normalize $\phi_0 = \sqrt{2} \kappa_D \phi_{\rm c.n.}$, and since $\kappa_D$ is kept fixed we simply get up to a numerical constant:
\ea{
&\kappa_D \left (2 - \sum_{i=1}^{n}  k_i^\mu \frac{\partial}{\partial k_{i}^\mu} Ê\right )
M_n 
 \sim
\frac{d}{d \phi_{\rm c.n.}} M_n \, 
\, .
}
Thus in both cases we find that, up to numerical constants, the $\mathcal{O}(q^0)$ term of the soft-dilaton amplitude is simply given by the derivative of the lower-point amplitude with respect to the vacuum expectation value.

Before we leave the term of $\mathcal{O}(q^0)$, let us conclude
with a more intuitive argument for the kinematically invariant terms of $\mathcal{O}(q^0)$. 
In the case of a NG dilaton,
 all dimensional factors are rescaled by a scale transformation, while in string theory
 one rescales  the factor $\frac{1}{\alpha'}$ in the front of  Eq. (\ref{Mndimana}) without
 rescaling $\kappa_D$. That is the reason why in one case one gets
$D - n \frac{D-2}{2}$, while in the other case one gets $2$.
   
Finally, the terms of $\mathcal{O}(q^1)$ are equivalent up to a single piece; 
in the case of the NG dilaton, Eq.~(\ref{finalbehavior}),
there is a term with a single derivative, which is not present in the case of the gravity dilaton.
As mentioned, however, they both contain the operator related to special conformal space-time translations.

It would be interesting to extend our considerations
to the loop diagrams of both dilatons. In string theory, the
dilaton stays massless to any order of perturbation theory, while 
in field theory, the dilaton, in general, gets a mass because conformal invariance is
explicitly broken in the quantum theory. (For a perturbatively controllable example, see for instance Refs.~\cite{Antipin:2011aa,Antipin:2012sm}.) There are, however, theories such as
${\cal{N}}=4$ super Yang-Mills on the Coulomb branch that are not plagued by 
a conformal anomaly. In these theories, it would be especially compelling to investigate 
the extent of agreement with our soft theorem at the quantum level.

\vspace{-5mm}
\subsection*{Acknowledgments} \vspace{-3mm}
We thank Massimo Bianchi, Rutger Boels, Andrea Guerrieri, Henrik Johansson,
and Congkao Wen for many useful discussions.


\end{document}